\documentclass[%
superscriptaddress,
 amsmath,amssymb,
 aps,
 reprint,
prb,
]{revtex4-2}

\usepackage{graphicx}
\usepackage{dcolumn}
\usepackage{bm}
\usepackage{physics}
\usepackage{siunitx}
\usepackage[normalem]{ulem}
\usepackage[caption=false]{subfig}

\usepackage[colorlinks=true, citecolor=blue, linkcolor=blue]{hyperref}
\usepackage[capitalize]{cleveref}   
\newcommand{\proptoinverse}{\mathrel{\mskip1mu\reflectbox{$\propto$}\mskip-1mu}}

\begin{document}

\title{Dynamically Dressed States of a Quantum Four-Level System}

\author{Carolin Calcagno}
 \email{Corresponding author: carolin.calcagno@tum.de}
 \affiliation{Walter Schottky Institut, Department of Electrical and Computer Engineering, and MCQST, Technische Universität München, 85748 Garching, Germany}
\author{Friedrich Sbresny}
 \affiliation{Walter Schottky Institut, Department of Electrical and Computer Engineering, and MCQST, Technische Universität München, 85748 Garching, Germany}
\author{Thomas K. Bracht}
 \affiliation{Institut für Festkörpertheorie, Universität Münster, 48149 Münster, Germany}
 \affiliation{Condensed Matter Theory, TU Dortmund, 44221 Dortmund, Germany}
\author{Sang Kyu Kim}
 \affiliation{Walter Schottky Institut, Department of Electrical and Computer Engineering, and MCQST, Technische Universität München, 85748 Garching, Germany}
\author{Eduardo Zubizarreta Casalengua}
 \affiliation{Walter Schottky Institut, Department of Electrical and Computer Engineering, and MCQST, Technische Universität München, 85748 Garching, Germany}
\author{Katarina Boos}
 \affiliation{Walter Schottky Institut, Department of Electrical and Computer Engineering, and MCQST, Technische Universität München, 85748 Garching, Germany}
\author{William Rauhaus}
 \affiliation{Walter Schottky Institut, Department of Electrical and Computer Engineering, and MCQST, Technische Universität München, 85748 Garching, Germany}
\author{Hubert Riedl}
 \affiliation{Walter Schottky Institut, TUM School of Natural Sciences, and MCQST, Technische Universität München, 85748 Garching, Germany}
\author{Jonathan J. Finley}
 \affiliation{Walter Schottky Institut, TUM School of Natural Sciences, and MCQST, Technische Universität München, 85748 Garching, Germany}
\author{Doris E. Reiter}
 \affiliation{Condensed Matter Theory, TU Dortmund, 44221 Dortmund, Germany}
\author{Kai Müller}
 \affiliation{Walter Schottky Institut, Department of Electrical and Computer Engineering, and MCQST, Technische Universität München, 85748 Garching, Germany}

\date{\today}

\begin{abstract}
    In this work, we experimentally and theoretically study the dressed-state emission of the biexciton-exciton cascade in a semiconductor quantum dot under pulsed, resonant, two-photon excitation.
    Building on the well-characterized steady-state dressed emission of the four-level system, we examine its dynamic counterpart under pulsed, resonant excitation, addressing both experimental observations and theoretical modeling.
    Here we report several sidebands emerging from the biexciton-to-exciton transition, whose number and spectral width depend on the excitation pulse duration and the effective pulse area, while no sidebands emerge from the exciton-to-ground-state transition.
    Since the biexciton state population follows a nonlinear pulse area function, sidebands with a small spectral nonlinearity result. 
    Detuning- and time-dependent measurements provide deeper insight into the emission properties of the dressed states. 
    They show that side peak emission only occurs in the presence of the excitation pulse. Moreover, when the system is excited by a Gaussian-shaped laser pulse, side peak emission takes place sequentially.
\end{abstract}

\maketitle

\section{Introduction}
\label{sec:introduction}
Investigating the resonant interaction between few-level quantum systems and light fields enables fundamental demonstrations of quantum optics \cite{glauber1963coherent&incoherentStates, mollow1969, kimble1977photonAntibunching, hongOuMandel1987}. 
Moreover, it is essential for the generation of non-classical states of light for applications in photonic quantum technologies \cite{obrien2009}. 
The coherent coupling of light and matter results in the formation of new eigenstates, the dressed states \cite{cCohen-Tannoudji_1977}.
For a two-level system resonantly driven by a strong continuous-wave (cw) light field, the resulting emission spectrum is the well-known Mollow triplet \cite{mollow1969}.
For pulsed excitation, the resulting dynamically dressed states have been predicted theoretically already in the 1980s \cite{KRzazewski_1984, florjanczyk1985, lewenstein1986, buffa1988, rogers1991} and observed experimentally in the 2000s for ensembles in microplasmas \cite{compton2009microplasma, compton2011microplasma}.
Just recently, Boos \textit{et al.} \cite{boos2024signatures} and Liu \textit{et al.} \cite{liu2024dynamic} provided the first experimental demonstration of these dynamically dressed states using excitonic transitions in semiconductor quantum dots (QDs) as a two-level system.
Semiconductor QDs are particularly well suited for such experiments due to their strong oscillator strength and nearly transform-limited linewidth \cite{Kuhlmann2015transformLimited, heindel2023QDreview}. 
Importantly, QDs can not only be used as prototypical two-level systems but also as four-level systems forming a two-path cascade, given by the biexciton-exciton cascade of uncharged QDs \cite{brunner1994}. 
For applications, this is promising for the deterministic generation of entangled photon pairs \cite{benson2000entangledPhotons, Lodahl2018quantumNetworks, trotta2021entangledPhotons, Rota2024entangledPhotons}. 
To this end, the excited state of the two-path cascade can be deterministically prepared via an effective two-photon transition \cite{stufler2006TPE,machnikowski2008twophoton}.

Here, we study the dynamic emission spectrum for the pulsed two-photon excitation (TPE) of the biexciton state in a semiconductor QD.
We observe the emergence of multiple side peaks from the biexciton-exciton transition, similar to those in a two-level system \cite{moelbjerg2012resonance, boos2024signatures, liu2024dynamic, kaspari2024unveiling}, while no side peaks are observed for the exciton-to-ground-state transition due to the temporal mismatch of the two emission lines in the cascaded decay.
This unilateral multi-peak emission spectrum can be explained by extending the cw dressed-state picture of the biexciton state to the dynamic emission spectrum under pulsed TPE.
We experimentally and theoretically characterize the multi-peak emission spectrum as a function of pulse duration, to develop a comprehensive understanding of the complete picture of resonant TPE.
Furthermore, we investigate how the pulse duration affects TPE Rabi rotations and establish a direct link between the appearance of the side peaks and the excited state population of the biexciton.
To confirm that the side peak emission occurs during the presence of the pulse, we study the behavior of the side peaks when the system is subjected to a finite detuning from the two-photon resonance and investigate the time-dependent evolution of the side peaks, both experimentally and theoretically.

\section{Biexciton System Dressing}
\label{sec:biexcitonSystemDressing}
\begin{figure}[ht]
    \centering
    \includegraphics{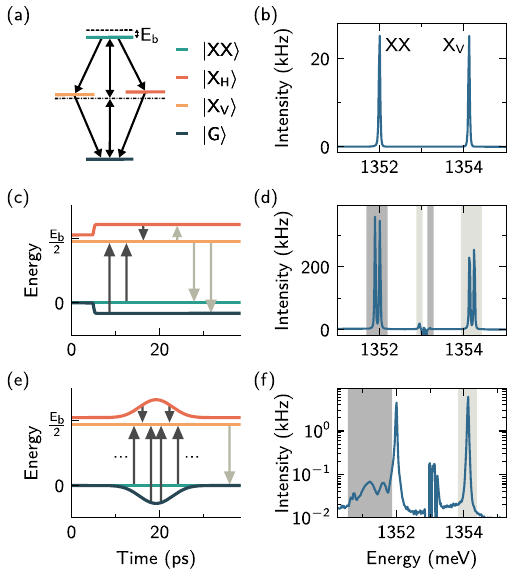}
    \caption{
    (a) Level scheme of the biexciton-exciton cascade. Double-sided arrows indicate the laser field. Arrows pointing downward indicate emission by radiative decay. 
    We excite the system with horizontally polarized laser light and collect only vertically polarized photons in a cross-polarized RF setup.
    (b) Measured emission spectrum after TPE of the biexciton. The biexciton and exciton emission energies are separated by the biexciton binding energy $\mathrm{E_b}$. 
    (c) Energy diagram of the biexciton-exciton cascade in the rotating frame of the excitation laser. At $\mathrm{t=\SI{5}{\ps}}$, the system is excited by a strong cw laser, resulting in the dressing and mixing of states. There are six allowed transitions, indicated by the arrows, when the system is driven on one polarization branch while detecting the other. Dark gray arrows indicate transitions from $\mathrm{\ket{XX} \rightarrow \ket{X_V}}$, light gray arrows indicate transitions from $\mathrm{\ket{X_V} \rightarrow \ket{G}}$.
    (d) Measured dressed emission spectrum of the biexciton-exciton cascade under strong cw excitation. Six optical transitions (dark and light gray background areas) and the imperfectly suppressed excitation laser are observed.
    (e) Dressed energy diagram of the biexciton-exciton cascade under pulsed excitation. The dressing of the states occurs only during the presence of the pulse, varies in time and follows the temporal shape of the excitation pulse. 
    Due to a time-dependent population of the excited states a modulated emission probability exists.
    (f) Measured dressed emission spectrum of the biexciton-exciton cascade under strong pulsed excitation in a semi-logarithmic scale. A one-sided multi-peak structure red-detuned from the biexciton transition appears.
    }
    \label{fig:theory}
\end{figure}
We describe the studied InGaAs QD as a four-level system as depicted in \cref{fig:theory}(a). The model consists of the ground state $\mathrm{\ket{G}}$, the two exciton states $\mathrm{\ket{X_V}}$ and $\mathrm{\ket{X_H}}$, which are energetically split by a small fine structure splitting (FSS) \cite{bayer2002finestructure}, and the biexciton state $\mathrm{\ket{XX}}$, which due to Coulomb interactions is energetically detuned from twice the exciton energy by the biexciton binding energy $\mathrm{E_b}$.
Transitions from the ground state to the biexciton state can be driven by a two-photon resonant, linearly polarized laser.
For this process, the laser is tuned to half the biexciton energy, represented by the dash-dotted line.
For InGaAs QDs this energy is usually detuned from the single exciton transitions by up to a few milli-electron volts.
In this basis, the decay of the biexciton leads to the emission of either a horizontally ($\mathrm{H}$) or vertically ($\mathrm{V}$) polarized photon, leaving the system in either the exciton state $\mathrm{\ket{X_H}}$ or $\mathrm{\ket{X_V}}$, respectively. The decay of these exciton states yields a second photon of the same polarization.
For a simpler theoretical model and experimental study of the spectrum of emitted photons, we focus our attention on the case where the excitation laser is horizontally polarized, while only the vertically polarized emission is studied. 
A typical experimentally measured emission spectrum of the quantum four-level system is shown in \cref{fig:theory}(b). There the $\mathrm{\ket{XX}}$ and $\mathrm{\ket{X_V}}$ emission lines are energetically separated by $\mathrm{E_b}$.

The QD and its interaction with the light field can be described using the Hamiltonian \cite{seidelmann2022twophoton} in the frame rotating with the laser frequency
\begin{align}
\begin{split}
    \mathrm{H} &= \mathrm{(\Delta + \frac{\delta}{2}) \ket{X_H}\bra{X_H} + (\Delta - \frac{\delta}{2}) \ket{X_V}\bra{X_V}} \\
    & + \mathrm{(2\Delta- E_b) \ket{XX}\bra{XX} - \frac{\hbar}{2}\Omega(t)(\sigma_L + \sigma_L^{\dagger})}
    \,,  
    \label{eq:hamiltonian_4ls}
\end{split}
\end{align}
where $\mathrm{\Delta = \hbar\omega_X-\hbar\omega_L}$, $\mathrm{\delta}$ is the FSS, $\mathrm{\Omega(t)}$ describes the time-dependent interaction of the system with the laser field, and where the laser polarization enters using
\begin{equation}
    \mathrm{\sigma_L} = \mathrm{\alpha_H \sigma_H + \alpha_V\sigma_V}
    \,.
\end{equation}
The parameters $\mathrm{\alpha_{H/V}}$ are the components of the laser with horizontal or vertical polarization and the transition operators read
\begin{equation}
    \mathrm{\sigma_{H/V}} = \mathrm{\ket{G}\bra{X_{H/V}} + \ket{X_{H/V}}\bra{XX}}
    \,.
\end{equation}
Diagonalizing the Hamiltonian in \cref{eq:hamiltonian_4ls} yields four eigenenergies corresponding to the dressed states (see \cref{app:theory} for a detailed theoretical description).

Exciting the system with a two-photon resonant cw laser results in the dressing of the states \cite{sánchezMuñoz_2015, hargart2016dressingX0andXX, pelle2016nonlinearDressedStates}. 
In \cref{fig:theory}(c), which depicts the energy diagram of the four-level system in the frame rotating with the laser frequency, this shows as an energy shift of the states when the cw laser is turned on at approximately $\mathrm{t=\SI{5}{ps}}$. 
Before excitation, the states $\mathrm{\ket{G}}$ and $\mathrm{\ket{XX}}$ at $\mathrm{E = \SI{0}{meV}}$ are degenerate due to the choice of the rotating frame, while the exciton states $\mathrm{\ket{X_V}}$ and $\mathrm{\ket{X_H}}$ are energetically located at half the biexciton binding energy.
Moreover, the two exciton states are separated by a comparably small FSS. 
With the light field present at $\mathrm{t > \SI{5}{\ps}}$, the interaction of the states with the laser field causes not only an energy shift but also a mixing of the states, given by the entries of the corresponding eigenvectors that form the dressed states.
Because the laser is horizontally polarized, the vertically polarized subsystem (orange line) experiences no mixing or energy shift. 
The mixing of the three other states (red, light green, dark green) allows transitions from the dressed states to $\mathrm{\ket{X_V}}$ and vice versa, represented by the dark and light gray arrows.
The dark gray arrows indicate the $\mathrm{\ket{XX} \rightarrow \ket{X_V}}$ dressed-state transitions, while the light gray arrows indicate the $\mathrm{\ket{X_V} \rightarrow \ket{G}}$ dressed-state transitions. An upward-pointing arrow denotes a negative energy shift and a downward-pointing arrow denotes a positive energy shift, both with respect to the laser frequency.
\cref{fig:theory}(d) displays the measured emission spectrum resulting from the cw dressing of the four-level system. It shows six emission peaks corresponding to the six allowed transitions represented by the dark and light gray arrows in \cref{fig:theory}(c). 
The dark gray background areas mark the $\mathrm{\ket{XX} \rightarrow \ket{X_V}}$ dressed-state transitions, while the light gray background areas mark the $\mathrm{\ket{X_V} \rightarrow \ket{G}}$ dressed-state transitions, respectively. 
The two peaks next to the laser energy are barely visible in the experiment due to comparatively lower intensities and only small spectral separation from the imperfectly suppressed excitation laser centered between the biexciton and exciton emission lines.

Under pulsed excitation, unlike for cw excitation, the dressing occurs only during the presence of the pulse and varies in time, following the temporal shape of the electric field of the pulse. 
For a Gaussian-shaped laser pulse this is illustrated in \cref{fig:theory}(e) in the energy diagram of the biexciton-exciton cascade depicting the dynamical dressing of the states.
During the finite pulse, which is about ten times shorter than the biexciton and exciton emission lifetimes, especially the biexciton population already starts to decay.
In \cref{fig:theory}(e) this is illustrated by the dark gray arrows which represent dressed-state transitions from $\mathrm{\ket{XX} \rightarrow \ket{X_V}}$.
In a time-integrated emission spectrum, the dressed-state decay leads to the appearance of side peaks in the spectrum. 
This can be observed in \cref{fig:theory}(f), where red-detuned from the biexciton emission energy (dark gray background area), three peaks of intensity lower than the biexciton emission line appear.
The total number of side peaks appearing in the spectrum depends on the number of Rabi cycles during the pulse, i.e., the time-dependent population of the excited state, and thus on the area of the pulse, with a higher pulse area giving rise to more peaks in the spectrum \cite{KRzazewski_1984, florjanczyk1985, lewenstein1986, moelbjerg2012resonance}.
The underlying physics of the appearance of the side peaks under pulsed excitation has been intensively studied for the two-level system and can be described via two different approaches.
One approach uses perturbed free induction decay \cite{kaspari2024unveiling}, while the other uses an approximate phase-matching model and interference of emission occurring at different times during the interaction with the laser pulse \cite{florjanczyk1985, moelbjerg2012resonance}. 
The second approach allows for an analytic approximation of the temporal emergence of the side peaks and suggests their spectral symmetry under resonant excitation.
The captured physics of the emergence of side peaks through pulsed excitation of a two-level system is similar for a four-level system. 
However, no analytic approximation exists and the problem has to be modeled numerically.
See \cref{subapp:theoryTimeDependentOccurrence} for the numerical approximation of the phase-matching model and how it translates to the dynamical dressing of the biexciton state.
Upon revisiting \cref{fig:theory}(f), a striking observation is that blue detuned from the exciton emission line, there is no multi-peak structure, i.e., the symmetry is broken.
This can be explained by the pulse duration which is short compared to the biexciton lifetime. 
When the exciton state starts to be significantly populated due to the cascaded emission process the dressing of the states has already disappeared.
Therefore, the multi-peak structure during the $\mathrm{\ket{X_V} \rightarrow \ket{G}}$ decay is strongly quenched and is not observed in \cref{fig:theory}(f) with the experimentally available signal-to-noise ratio.
In \cref{fig:theory}(e) this phenomenon is illustrated by a single light gray arrow, denoting the exciton-to-ground-state transition when the state dressing is long gone.
Note that the theory predicts additional side peaks blue-detuned from the driving laser energy, represented by the dark gray, downward-pointing arrows in \cref{fig:theory}(e). 
These side peaks are of such weak intensity that they cannot be resolved in the experiment.

In this work, we perform theoretical simulations using the Hamiltonian given in \cref{eq:hamiltonian_4ls} and consider coupling to LA phonons using a numerically exact PT-MPO method \cite{cygorek2022simulation,cygorek2024sublinear,cygorek2024ace}. We retrieve the spectra from the first-order optical coherence, $\mathrm{G^{(1)}(t,\tau) = \langle\sigma^{\dagger}_V(t+\tau)\sigma_V(t)\rangle}$, as detailed in Refs.~\cite{moelbjerg2012resonance, boos2024signatures} and \cref{app:theory}.

\section{Dynamical Dressing of the Biexciton System}
\label{sec:dynamicalDressingOfTheBiexcitonSystem}
Experimentally, Gaussian pulses with variable pulse duration and tunable wavelength are generated via a custom-built $\mathrm{4f}$ pulse-shaping setup in the same polarization configuration as described above.
To efficiently suppress the horizontally polarized laser, the QD is studied in a cross-polarized resonance fluorescence setup.
Further details regarding sample structure, measurement setup and optical characterization can be found in \cref{app:experimentalDetails}.
\begin{figure}
    \centering
    \includegraphics{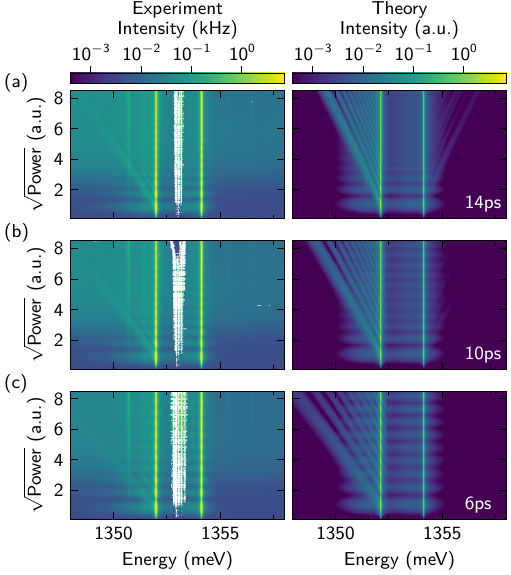}
    \caption{
    Logarithmically plotted intensity of measured and simulated power-dependent emission spectra.
    Rabi oscillations between the $\mathrm{\ket{G} \leftrightarrow \ket{XX}}$ transition are driven by pulsed resonant TPE for pulse durations of (a) \SI{14}{\pico\s}, (b) \SI{10}{\pico\s}, and (c) \SI{6}{\pico\s}.
    In addition to the biexciton and exciton emission lines, multiple sidebands emerge from the biexciton transition. As the excitation power increases, an increasing amount of side peaks appear which shift towards lower energies, resulting in greater detuning from the biexciton emission line. 
    The strongly quenched side peak emission emerging from the exciton-to-ground-state transition is only visible in the theoretical calculations with a pulse duration of (a) \SI{14}{\pico\s}.
    With decreasing excitation pulse duration the dressed part of the spectrum broadens, and the number of side peaks decreases within a given power range.
    }
    \label{fig:colormap}
\end{figure}

To experimentally study the complete emission spectrum under pulsed excitation, time-integrated spectra are collected as shown in \cref{fig:colormap}. The plots display the logarithmic emission intensity versus both the emission energy and the square root of the excitation power for different pulse durations, for both experiment (left) and theory (right).
In \cref{fig:colormap}(a), short Gaussian pulses with an intensity full width at half maximum (FWHM) of \SI{14}{ps} drive Rabi rotations of the $\mathrm{\ket{G} \leftrightarrow \ket{XX}}$ transition via two-photon resonant excitation.
The biexciton and exciton emission lines emitting at energies of \SI{1352.0}{\meV} and \SI{1354.1}{\meV}, respectively, are energetically separated by the binding energy of \SI{2.1}{meV}. 
As the excitation power increases, oscillating intensities of both emission lines arise from the driven Rabi rotation.
Between these emission lines, the spectrally separated excitation laser is visible in the experimental data as a broad vertical white bar due to imperfect cross-polarized suppression and the used on-off measurement technique (see \cref{subapp:On/Off} for details). This combination can result in an integrated count rate outside the color scaling of the logarithmic emission intensity.

In the logarithmic scaling, we observe prominently multiple sidebands emerging from the biexciton transition energy.
Centered at each maximum of occupation a new side peak appears, and gradually shifts to lower energies as the driving strength increases.
This discrete occurrence results in a very specific number of sidebands, which is determined by the driving strength and pulse width -- that is, by the effective pulse area, which is here defined as the number of two-photon Rabi rotations of the biexciton state population.
Because the dressing of the states occurs only during the presence of the pulse, and since the excitation pulse with a pulse width of $\mathrm{\SI{14}{\ps}}$ is short compared to the biexciton lifetime of $\mathrm{\SI{157}{\ps}}$ (see \cref{subapp:Lifetime_XX_X}), the emission of the exciton-to-ground-state transition during the pulse duration is strongly suppressed by the temporal delay introduced by the cascaded nature of the four-level system.
This results in the absence of the multi-peak structure for the exciton emission line for the available signal-to-noise in the experiment.
However, there exists a finite probability that both the biexciton and the exciton state quickly decay while the pulse is still active. This can be seen in the theoretical calculations where we observe faint sidebands evolving from the exciton emission line for a pulse duration of \SI{14}{ps} in \cref{fig:colormap}(a).
Additionally, we observe a slight mismatch between the experiment and the theory regarding the spectral position of the side peaks which we attribute to the laser pulses in the experiment not being perfectly Gaussian shaped.

In general, the intensity of the observed side peaks is low compared to the main emission lines of the biexciton and exciton decays. Moreover, high background luminescence and the phonon sideband further reduce the visibility.
The phonon sideband, observed both experimentally and theoretically, appears as broad horizontal emission lines that follow the intensity oscillation of the Rabi rotations. It originates from lattice relaxation and the emission of a phonon, resulting in the emission of a photon with energy lower than the exciton or biexciton transition energy \cite{krummheuer2002theory, reiter2014role, roy2015quantum, phononJune2017}.
The background luminescence, which is only observed in the experiment, increases with increasing excitation power, as seen by the transition from dark blue to light blue in the background intensity. Its origin remains uncertain, however, we assume it to be due to background luminescence from impurities and charge traps within the semiconductor host matrix.
Additionally, the experimental data exhibit vertical stripes, which we attribute either to non-resonantly driven states of nearby QDs or to additional charge states of the studied quantum dot that appear under strong pulsed excitation.

The emergence and position of the side peaks depend solely on the pulse area and the resulting dressing of the states. 
To support this statement, we study emission spectra for different pulse durations. \cref{fig:colormap}(b) and \ref{fig:colormap}(c) show the emission spectra for pulse durations of $\mathrm{\tau=\SI{10}{\ps}}$ and $\mathrm{\tau=\SI{6}{\ps}}$, respectively.
By comparing panels (a--c) of \cref{fig:colormap}, two main observations can be made. 
First, with decreasing pulse duration, the individual side peaks shift to lower energies, i.e., the dressing increases for similar pulse areas. 
Second, shorter pulses lead to fewer Rabi oscillation cycles within a given power range and thus to a reduced absolute number of emerging side peaks. 
To understand these dynamics, we take a closer look at the two-photon Rabi rotations for different pulse durations and their relation to the dressing strength.

\section{Pulse Length Analysis}
\label{sec:pulseLengthAnalysis}
\begin{figure}
    \centering
    \includegraphics{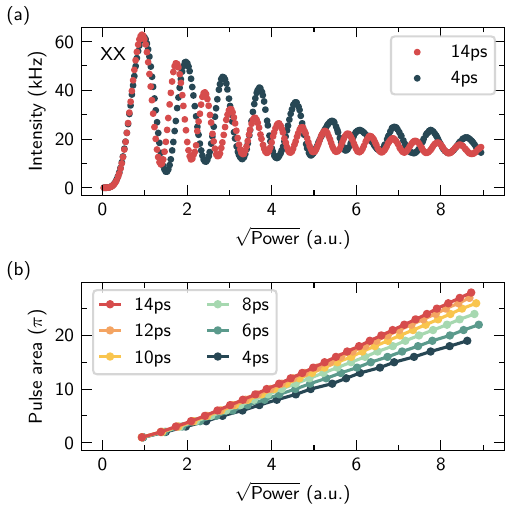}
    \caption{
    (a) Rabi rotations of the $\mathrm{\ket{G} \leftrightarrow \ket{XX}}$ transition driven by resonant TPE with Gaussian pulses of duration of $\mathrm{\SI{14}{\ps}}$ (red) and $\mathrm{\SI{4}{\ps}}$ (dark teal).
    The first $\mathrm{\pi}$ rotation requires the same average excitation power regardless of the pulse length. With increasing excitation power, the Rabi rotations begin to diverge. For both Rabi rotations, a nonlinear behavior with the square root of the excitation power is observed.
    (b) Pulse areas of Rabi rotations for six different pulse durations ranging from $\mathrm{\SI{14}{\ps}}$ down to $\mathrm{\SI{4}{\ps}}$ are plotted as a function of the square root power.
    For the $\mathrm{\SI{4}{\ps}}$ pulse, the pulse area is almost linear with the square root of the excitation power. For longer pulses, the nonlinear and pulse length dependent nature of TPE becomes visible in an upward slope.
    All curves show the same average excitation power for the first $\mathrm{\pi}$ rotation.
    With decreasing pulse duration, the number of $\mathrm{\pi}$ rotations decreases for the same power range.
    }
    \label{fig:pulseArea}
\end{figure}
In the experiment, we manipulate the biexciton state population by changing the effective pulse area. We do this by increasing the average excitation laser power while leaving the pulse duration constant.
In \cref{fig:pulseArea}(a) two Rabi rotations of the $\mathrm{\ket{G} \leftrightarrow \ket{XX}}$ transition are driven by resonant TPE with both a $\mathrm{\tau = \SI{14}{\ps}}$ (red) and a $\mathrm{\tau = \SI{4}{\ps}}$ (dark teal) long pulse, respectively. The integrated emission intensity of the biexciton is plotted as a function of the square root of the average excitation power.
For low excitation powers, we observe that up to the first complete inversion of the biexciton state the same excitation power is required for both pulse durations. 
From this observation, we find that for low driving the excitation power needed to achieve the same population inversion of the biexciton state is independent of the pulse length.
Moreover, when we convert the theoretical model from an effective pulse area describing the biexciton occupation in the low power regime by Stufler \textit{et al.} \cite{stufler2006TPE} into the power domain (see \cref{app:pulsearea}), we observe that the TPE Rabi oscillations depend linear on the excitation power.
Relating this to the dressed-state picture, we find that having the same average excitation power to achieve the first inversion of the biexciton state with decreasing pulse lengths results in an increased peak power for shorter pulses and thus a higher field intensity.
As the excitation power increases, the two Rabi rotations begin to diverge.
The Rabi rotations driven by the $\mathrm{\SI{14}{\ps}}$ pulse start to oscillate faster with increasing power and show rotations up to $\mathrm{28\pi}$. Meanwhile, in the same power range the Rabi rotations driven by the $\mathrm{\SI{4}{\ps}}$ pulse show rotations up to only $\mathrm{19\pi}$.
With increasing excitation power and thus increasing area of the pulse the two-photon Rabi rotations gradually convert to a square root dependence on power, as it is known from Rabi rotations of a resonantly driven two-level system. 
This convergence from linear to square-root power dependence is slow and depends on the pulse duration $\mathrm{\tau}$ \cite{stufler2006TPE}.
Thus, as the excitation pulse length decreases, the low driving limit with linear power dependence lasts shorter, resulting in a faster convergence to the square root power dependence and thus fewer Rabi rotations.
This trend becomes even more apparent when looking at \cref{fig:pulseArea}(b). 
There we plot the effective pulse areas ($\mathrm{\pi}$), denoting the maxima and minima biexciton occupation, for a total of six Rabi rotations with pulse durations ranging from $\mathrm{\SI{14}{\ps}}$ down to $\mathrm{\SI{4}{\ps}}$ as a function of the square root of the average excitation power.
While all curves shown have the same average power required for the initial $\mathrm{\pi}$ rotation, as the power increases, they start to diverge. With increasing excitation pulse lengths we observe a gradually increasing upward slope, while for short pulses the curves approach an almost linear dependence on the square root of the excitation power.
At the same time, with decreasing excitation pulse length, the number of $\mathrm{\pi}$ rotations gradually decreases within the same power range.
These observations explain both findings when comparing panels (a--c) in \cref{fig:colormap}.
As the excitation pulse length decreases, the number of emerging side peaks also decreases, because they are directly related to the number of Rabi cycles that can be realized for a given power.
Second, for the same average power shorter pulses have a larger maximum field intensity which causes a stronger instantaneous state dressing.
Because all Rabi rotations have their first population inversion at the same average power, this results in a larger detuning of the side peaks from the main emission line for shorter pulse durations.
This trend is further enhanced by the nonlinear pulse-length dependent behavior of the two-photon Rabi rotations, because shorter pulses require a larger average power for any multiple $\mathrm{\pi}$ rotation, which can be clearly seen in \cref{fig:pulseArea}(b).

Given that the side peaks are determined by the excited state population of the biexciton, the spectral movement of the side peaks to larger detunings follows the curves shown in \cref{fig:pulseArea}(b), where a small degree of nonlinearity exists. This is most clearly visible in \cref{fig:colormap}(a), where the system is excited with a long laser pulse of length \SI{14}{\ps}.

\section{Detuning Dependence}
\label{sec:detuningDependence}
\begin{figure}
    \centering
    \includegraphics{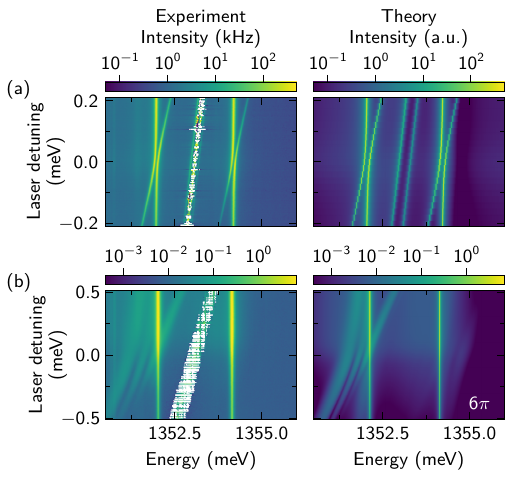}
    \caption{
    Measured and simulated emission spectra under detuned cw and pulsed excitation.
    (a) A strong cw laser is tuned across the TPE resonance over a range of $\mathrm{\pm \SI{0.21}{\milli\eV}}$. 
    Six emission lines appear. The dressed part of the spectrum follows the detuning of the excitation laser, while at resonance the shifting lines anticross with the non-shifting lines.
    (b) A pulsed excitation laser with a pulse length of \SI{14}{\pico\s} and a pulse area of $\mathrm{6\pi}$ is tuned across the TPE resonance by $\mathrm{\pm \SI{0.51}{\milli\eV}}$.
    Under pulsed excitation, the biexciton and exciton transition lines maintain their transition energies.
    For $\mathrm{\Delta < 0}$, the side peaks follow the detuning of the excitation laser. For $\mathrm{\Delta >0}$ the side peaks asymptotically approach the biexciton emission line, with the innermost side peak continuing to follow the laser detuning. 
    }
    \label{fig:detunedMollow}
\end{figure}
To further investigate the time-integrated dynamic emission spectrum, we study additional emission spectra with finite detuning from the two-photon resonance under both cw and pulsed excitation, as presented in \cref{fig:detunedMollow}. 
As before, the experimental data are shown in the left panel and the corresponding simulations are shown in the right panel.
\cref{fig:detunedMollow}(a) depicts the dressing of the states under finite detuning for cw excitation. For these measurements, the excitation laser is scanned by $\mathrm{\pm\SI{0.21}{\meV}}$ across the two-photon resonance while maintaining a constant high power of $\mathrm{\SI{500}{\micro\W}}$.
As predicted by theory, in total six emission lines are visible in both experiment and theory, while due to imperfect laser suppression the residual laser light is visible in the experimental data.
The dressed part of the spectrum follows the detuning of the excitation laser, causing a shift in the emission energies. 
At resonance, the shifting lines of the spectrum anticross with the non-shifting lines.
\cref{fig:detunedMollow}(b) shows the spectrum for pulsed excitation, where the \SI{14}{\ps} long excitation pulse is scanned by $\mathrm{\pm\SI{0.51}{\meV}}$ across the two-photon resonance, with a constant power corresponding to a pulse area of $\mathrm{6\pi}$. 
In both experiment and theory, the emission spectra show the biexciton and exciton emission at their characteristic energies, with enhanced intensities for laser detunings $\mathrm{\Delta \geq 0}$ due to resonant ($\mathrm{\Delta=0}$) and phonon-assisted ($\mathrm{\Delta>0}$) excitation \cite{phononAssisted2014Per-Lennart, phononAssisted2015Quilter, phononAssisted2015Bounouar}.
Looking at the emission of the side peaks, we find that for negative laser detunings $\mathrm{\Delta<0}$, their emission energy follows the detuning and thus shifts to lower energies. In contrast, for positive laser detunings $\mathrm{\Delta>0}$, the side peaks asymptotically approach the main biexciton emission line, except for the innermost side peak, which continues to follow the laser detuning beyond the biexciton emission line to higher energies.
Interestingly, only the side peaks exhibit an energy shift following the laser detuning, while the biexciton and exciton emission happens at their characteristic energies. This finding strongly suggests that the side peak emission occurs during the pulse when the dressing is present, while most of the biexciton population decays via the cascade when there is no driving field present anymore.

\section{Temporal Behavior}
\label{sec:temporalBehavior}
\begin{figure}
    \centering
    \includegraphics{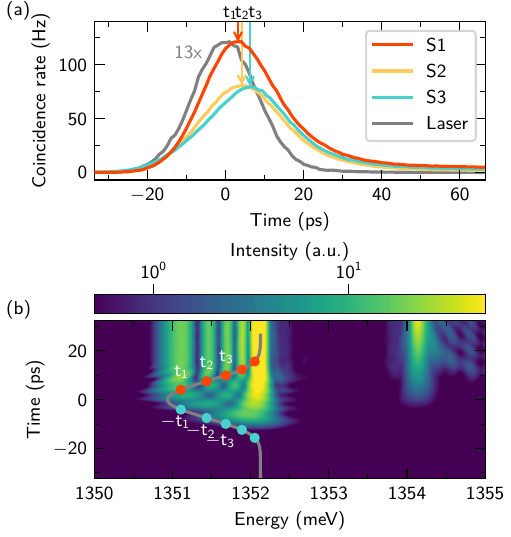}
    \caption{
    (a) Time-resolved measurements of the side peaks $\mathrm{S_1}$ (red), $\mathrm{S_2}$ (yellow), and $\mathrm{S_3}$ (blue). The measured temporal width of the side peaks is short compared to the biexciton and exciton emission lifetimes and follows the temporal profile of a Gaussian pulse with a small additional tail at the falling edge. The side peaks appear sequentially in time at $\mathrm{t_1}$ (red arrow), $\mathrm{t_2}$ (yellow arrow), and $\mathrm{t_3}$ (blue arrow).
    (b) Simulated time-integrated emission spectrum of the biexciton excited via TPE by a $\mathrm{\SI{14}{\ps}}$ Gaussian pulse (gray) with a pulse area of $\mathrm{10\pi}$, plotted as a function of time. Five side peaks (blue markers) start to detach from the biexciton emission line, while at $\mathrm{t_1}$, $\mathrm{t_2}$, and $\mathrm{t_3}$ (red markers) the side peaks $\mathrm{S_1}$, $\mathrm{S_2}$, and $\mathrm{S_3}$, respectively, start to gain strength and lock in place.
    }
    \label{fig:timeDependent}
\end{figure}
To better understand the temporal occurrence of the side peaks, we performed time-dependent measurements.
For this purpose, we correlated the photodiode signal of the pulsed laser with the emission of three individual side peaks.
To independently collect the specific emission energies, we used a custom-built spectral $\mathrm{4f}$-filter. The spectrally filtered side peak emission is detected using a superconducting nanowire single-photon detector, which together with the used time-to-digital converter has an overall timing jitter of $\mathrm{\sim \SI{17}{\ps}}$.
The results of these measurements are presented in \cref{fig:timeDependent}(a).
A Gaussian-shaped laser pulse (gray) with a FWHM of \SI{14}{\ps}, a pulse area of $\mathrm{10 \pi}$, and a maximum field intensity at $\mathrm{t = \SI{0}{\ps} }$ excites the biexciton-exciton cascade.
Note that in \cref{fig:timeDependent}(a) the excitation laser pulse is amplified by 13 times its measured intensity.
Upon excitation, the time-resolved emission of three individual side peaks, $\mathrm{S_1}$ (red), $\mathrm{S_2}$ (yellow), and $\mathrm{S_3}$ (blue), is measured.
The numbers 1, 2, and 3 refer to the side peaks that subsequently appear in the spectrum with increasing energy (\cref{fig:timeDependent}(b)).
We observe that the temporal width of the side peaks is significantly shorter than the biexciton ($\mathrm{\tau_{XX} = \SI{157}{\ps}}$) and exciton ($\mathrm{\tau_{X} = \SI{295}{\ps}}$) emission lifetimes (see \cref{subapp:Lifetime_XX_X}). Their temporal profiles resemble a Gaussian shape similar to the excitation pulse, with a barely visible tail at longer timescales. 
We attribute this tail to residual phonon-assisted emission at the spectrally filtered energies.
Based on these observations, the emission decay times of the side peaks are expected to be on the order of the excitation pulse length of $\mathrm{\tau = \SI{14}{\ps}}$. 
Due to the detector timing jitter of the order of the pulse duration, an exact determination of their temporal widths is unfeasible and only allows us to verify their short liveliness.
Interestingly, despite the limited timing resolution, we observe a temporal ordering of the emissions, with $\mathrm{S_1}$, $\mathrm{S_2}$ and $\mathrm{S_3}$ occurring in a sequential manner.
This finding has already been observed for the two-level system \cite{liu2024dynamic}. 
In \cref{fig:timeDependent}(a) the arrows mark the sequential times $\mathrm{t_1}$ (red), $\mathrm{t_2}$ (yellow) and $\mathrm{t_3}$ (blue) at which the time-dependent emission of the side peaks reaches its peak coincidence rate.

This observation can be explained by a theoretically calculated time-dependent spectrum, as defined in \cite{JHEberly_1980timeDpendentSpectrum}, and shown in \cref{fig:timeDependent}(b).
The plot displays the logarithmic emission intensity, which is integrated up to a point in time $\mathrm{t}$ and is plotted against both the time $\mathrm{t}$ and the emission energy.
Again, a single Gaussian pulse (gray) with a maximum field intensity at $\mathrm{t = \SI{0}{\ps} }$, a pulse duration of \SI{14}{\ps}, and a pulse area of $\mathrm{10 \pi}$ excites the biexciton-exciton cascade.
During the pulse, the biexciton state is populated and starts to emit with its lifetime of $\mathrm{\SI{157}{\ps}}$, which we observe by the dominant biexciton emission line at an energy of \SI{1352.0}{meV}. 
As the pulse evolves over time, five side peaks, one after the other, begin to detach from the main emission line, shift outwards to lower energies and begin to interfere with each other. Until, at a certain time $\mathrm{t_n}$ (red markers), each peak successively locks into its final spectral position. 
This behavior is consistent with the theory of Refs. \cite{florjanczyk1985, moelbjerg2012resonance}, which state that side peaks appear in a spectrum as a result of constructive interference from emission occurring at different times during the interaction with the pulse.
Thus, to generate a side peak in the spectrum using a temporally symmetric pulse, the pulse must have evolved until the point in time $\mathrm{t_n}$, where both $\mathrm{-t_n}$ (blue markers) and $\mathrm{t_n}$ (red markers) are included.
This confirms our finding that side peak emission occurs only during the presence of the excitation pulse and that a temporal ordering of the emergence of the side peaks in the spectrum exists.
Note that the 5\textsuperscript{th} side peak emits at the onset of the biexciton emission line and cannot be observed as a single peak.

Due to the finite picosecond resolution of our detectors, time-resolved measurement of the interference fringes at the rising edge of the pulse is not within the possibilities of our current experimental setup. However, we provide a simulation and discussion of these features in \cref{subapp:timeDependentIRF}.

\section{Conclusion}
\label{sec:conclusion}
In this work, we have experimentally investigated the spectral and temporal emission properties of the biexciton-exciton cascade under resonant TPE driven by finite Gaussian laser pulses. Our results are confirmed and extended by theoretical calculations.
Under pulsed excitation, dressing of the states occurs only during the presence of the excitation pulse and varies with time. 
For TPE of the biexciton state a unique unilateral side peak emission spectrum emerges, where side peaks evolve from the biexciton transition. In contrast, side peaks evolving from the exciton transition are strongly quenched due to the temporal delay introduced by the cascaded decay. This implies that side peak emission occurs during the pulse duration, which is confirmed by detuning and time-dependent photoluminescence measurements.
Moreover, we found that the number of side peaks, their detuning and spectral movement depend on a combination of both the electric field amplitude induced state dressing and the effective pulse area. 
Our findings contribute to a fundamental understanding of the complex time-dependent interaction of a laser pulse with the four-level system of a biexciton cascade in a quantum dot. Furthermore, they can be easily extended to any quantum multi-level system under resonant excitation. 
This enables future realizations of non-classical light generation such as highly correlated photons from dynamically dressed few-level systems \cite{2025arXivEdu} to explore leap-frog processes \cite{Valle_2013, Gonzalez-Tudela_2013} and complex correlated \cite{zubizarreta2024two} or entangled states of light.

\section*{Acknowledgments}
We gratefully acknowledge financial support from the Deutsche Forschungsgemeinschaft (DFG, German Research Foundation) via projects MU4215/4-1 (CNLG), INST 95/1220-1 (MQCL) and INST 95/1654-1 (PQET), Germany’s Excellence Strategy (MCQST, EXC-2111, 390814868), the Bavarian Ministry of Economic Affairs (StMWi) via project 6GQT and the German Federal Ministry of Education and Research via the project 6G-life.

\bibliography{Biexciton_dynamical_dressing}

\newpage
\onecolumngrid
\appendix

\section{Theory}
\label{app:theory}
In the theoretical model, we use a four-level system to describe the quantum dot, corresponding to the model used in Ref.~\cite{seidelmann2022twophoton}.
The Hamiltonian (not yet transformed to a rotating reference frame) describing the QD reads
\begin{equation}
    \mathrm{ H_0 = \left(\hbar\omega_X + \frac{\delta_{FSS}}{2}\right)\ket{X_H}\bra{X_H} + \left(\hbar\omega_X - \frac{\delta_{FSS}}{2}\right)\ket{X_V}\bra{X_V} + (2\hbar\omega_X-E_b)\ket{XX}\bra{XX} }
    \,.
\end{equation}
The interaction of the QD with the electric field of the laser is given by the interaction Hamiltonian in the dipole and rotating wave approximation
\begin{equation}
    \mathrm{ H_I = -\frac{\hbar}{2}(\Omega_H^*(t)\sigma_H + \Omega_H(t)\sigma_H^{\dagger}) - \frac{\hbar}{2}(\Omega_V^*(t)\sigma_V + \Omega_V(t)\sigma_V^{\dagger}) }
    \,,
\end{equation}
where the quantity $\mathrm{\Omega_{H/V}(t)}$ is proportional to the field of the $\mathrm{H/V}$-polarized component $\mathrm{\alpha_{H/V}}$ of the laser pulse, such that $\mathrm{\alpha_H^2 + \alpha_V^2 = 1}$ and $\mathrm{\Omega_{H/V}(t) = \alpha_{H/V}\Omega (t)}$. The laser field couples to the system using the polarization operators $\mathrm{\sigma_{H/V} = \ket{G}\bra{X_{H/V}} + \ket{X_{H/V}}\bra{XX}}$. 
In the following we consider Gaussian laser pulses of the form
\begin{equation}
    \mathrm{ \Omega(t) = \frac{\Theta}{\sqrt{2\pi}\tau_{field}}e^{-\frac{t^2}{2(\tau_{field})^2}}e^{-i\omega_L t} }
    \,,
\end{equation}
where $\mathrm{\omega_L}$ is the laser frequency and $\mathrm{\tau_{field}}$ is a measure of the pulse duration of the electric field. 
This pulse duration is related to the measured FWHM of the intensity autocorrelation function of the laser pulse by
\begin{equation}
    \mathrm{ \tau_{\mathrm{FWHM}}^{\mathrm{autoco}} = 2\sqrt{2\ln 2} \tau_{field} } = \sqrt{2}\tau
    \,, 
\end{equation}
where $\mathrm{\tau}$ denotes the pulse duration used in the experimental part of the main manuscript, which defines the intensity FWHM of a Gaussian laser pulse.

From the theoretical model, the time-dependent spectrum can be calculated using the multi-time correlation function $\mathrm{G^{(1)}(t,\tau) = \langle\sigma^{\dagger}(t+\tau)\sigma(t)\rangle}$, using \cite{moelbjerg2012resonance}
\begin{equation}
    \mathrm{ S(\omega,t) = \mathrm{Re}\left[\int_{-\infty}^{t}dt' \int_{-\infty}^{t-t'}d\tau \, G^{(1)}(t',\tau)e^{-i\omega\tau}\right] }
    \,.
\end{equation}
For $\mathrm{t\rightarrow\infty}$, this approaches the time-independent spectrum as it is measured as the integrated counts on the spectrometer. The $\mathrm{\sigma}$ operators used in the $\mathrm{G^{(1)}}$ function determine which part of the spectrum is calculated: for example, if $\mathrm{\sigma=\sigma_H}$ is used, only the horizontally polarized light is considered. The full spectrum (not resolved in polarization), could be obtained using $\mathrm{\sigma = \frac{1}{\sqrt{2}}(\sigma_H + \sigma_V)}$. In the main paper we always choose horizontally polarized pulses for excitation and calculate the spectrum of the emitted vertically polarized photons.

The dressed state picture used in the introduction of the main part of the manuscript is obtained by diagonalizing the Hamiltonian $\mathrm{H=H_0+H_I}$, transformed into a rotating reference frame rotating with the frequency of the two-photon resonant laser pulse.\\
This leads to four dressed state energies $\mathrm{E_i}$ (the eigenvalues obtained by diagonalizing the Hamiltonian for each point in time), which are shown under the action of the pulse in Fig.~1 of the main manuscript. Additionally, the bare states $\mathrm{\{\ket{G},\ket{X_H},\ket{X_V},\ket{XX}\}}$ mix to form the dressed states $\mathrm{\{\ket{\psi_1},\ket{\psi_2},\ket{\psi_3},\ket{\psi_4}\}}$, according to the entries $\mathrm{a_{ij}}$ of the obtained eigenvectors:
\begin{equation}
    \mathrm{ \ket{\psi_i} = \sum_{j=G,X_H,X_V,XX} a_{ij}\ket{j} }
    \,.
\end{equation}
This mixing leads to numerous possible transitions between the dressed states, which are discussed in the main part. For example, transitions from the biexciton to the exciton states transfer to transitions from all dressed states containing a finite contribution of the biexciton to all dressed states containing contributions of the exciton states. This includes ``transitions'' that start and end in the same dressed state if it is a mixture of biexciton and exciton. However, this does not show up in the cases we considered in the main part of the manuscript, since diagonalization under the action of a purely $\mathrm{\sigma_H}$-polarized laser pulse does not induce mixing of $\mathrm{X_V}$.

\section{Experimental Details}
\label{app:experimentalDetails}

\subsection{Sample Structure}
\label{subapp:sampleStructure}
The semiconductor quantum dot studied in this paper was grown by molecular beam epitaxy (MBE) using the Stranski-Krastanov growth process. By depositing four monolayers of InGaAs on GaAs, QDs are formed by strain release caused by a GaAs/InGaAs lattice mismatch. A distributed Bragg reflector (DBR), consisting of 17 alternating pairs of AlAs and GaAs layers, beneath the QD layer acts as a highly reflective mirror to enhance the collection efficiency of the emitted photons. To enable deterministic control of the QD charge state and to fine-tune their emission energies via the quantum confined Stark effect \cite{finley2004starkShift}, the QDs are embedded in a p-i-n-i-n diode structure. The n-doped interlayer allows the QDs to be charged close to zero bias by reducing the built-in electric field \cite{lobl2017pinin}.

\subsection{Optical Characterization}
\label{subapp:opticalCharacterization}
All measurements were performed in a closed-cycle cryostate with a base temperature of \SI{1.7}{\K}. 
To generate picosecond pulses with a Gaussian pulse shape, a femtosecond pulse from a Ti:Sapphire laser (Coherent Mira900f) is sent to a custom-built folded 4f pulse shaper.
The pulse shaper consists of a 1800 grooves per mm (g/mm) grating, a \SI{300}{\mm} focal length lens and an adjustable slit on a translation stage in the Fourier plane.
By varying the width and position of the slit in the Fourier plane, the temporal width and energy of the pulse can be tuned.
The pulse durations were determined by the full width at half maximum of the intensity autocorrelation and a subsequent deconvolution by a factor of $\mathrm{\sqrt{2}}$, assuming a Gaussian pulse shape.

A resonance fluorescence (RF) setup was used to efficiently filter the excitation laser from the quantum dot emission, with a cross-polarized filter in the detection path suppressing the linearly polarized excitation beam. 
We study the emission spectra using a Czerny--Turner spectrograph with a focal length of \SI{750}{\mm}, a 1200 g/mm grating, and a 2D back-illuminated, deep-depleted SiCCD camera attached to it.
To resolve the strongly suppressed dynamically dressed states, we maximize the signal-to-noise ratio by two measures: First we integrate for \SI{2.5}{\s} to exploit the full range of the 16bit DAC of the camera. Second, we average each spectrum three times to minimize the pixel-to-pixel noise.
To allow for a reasonable presentation of the contour plots with logarithmically plotted intensity an overall flat offset of \SI{4}{\Hz} was added to the emission spectra.
For the time-resolved measurements a superconducting nanowire single-photon detector (SNSPD) with a timing jitter of $\mathrm{\sim \SI{17}{\pico\s}}$ was used. To spectrally filter the individual peaks of the emission spectrum for the time-resolved measurements, an unfolded 4f filter with two \SI{1800}{g/\mm} blazed gratings and a bandwidth of $\mathrm{\sim\SI{0.23}{\meV}}$ was used.

\subsection{On/Off Measurement Technique}
\label{subapp:On/Off}
\begin{figure}
    \centering
    \includegraphics{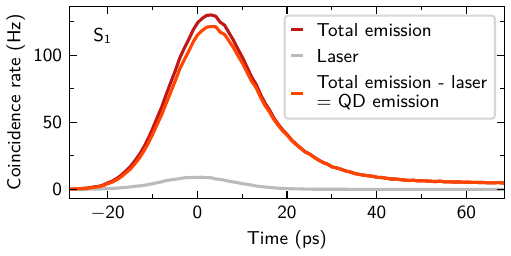}
    \caption{Time-resolved on/off measurement technique: In addition to the total emission (dark red), which includes the QD and laser emission, the imperfectly suppressed laser (gray) is measured with the QD electrically tuned out of resonance. To obtain only the QD emission (orange red), the laser (gray) is subtracted from the total emission (dark red).
    }
    \label{fig:OnOffOnOff}
\end{figure}
To account for imperfect laser suppression, on/off measurements were performed both for the spectral analysis and the time-resolved measurements.
For the spectral analysis, an on/off measurement refers to recording each emission spectrum alongside a corresponding reference spectrum, taken with identical averaging and integration time, where the QD is electrically tuned out of resonance while the laser is still shining on the sample.
This reference spectrum is then subtracted from the spectrum with the QD driven by the laser and electrically in resonance, i.e., where the photoluminescence of the QD is observed.
This measurement method, combined with the readout noise of the spectrometer, can result in negative intensity values at the energy of the laser pulse. 
This can be seen in \cref{fig:power-dependentColormap,fig:detunedColormap} and in all the emission spectra presented in the main manuscript in the form of white spectral points in the logarithmic color scale. 

For the time-resolved measurements, a corresponding reference measurement was again recorded with the QD electrically tuned out of resonance, while keeping the filter spectral position constant. This reference was then subtracted from the total coincidence rate as shown in \cref{fig:OnOffOnOff}. In the figure, the total emission of $\mathrm{S_1}$ is displayed in dark red, with the laser contribution (gray) leaking into the signal due to imperfect filtering.
The resulting subtracted coincidence rate, representing only the side peak emission, is shown in orange-red.
For all time-resolved measurements presented in the main manuscript, the subtracted coincidence rates (orange-red) were used.

\section{Power-Dependent Emission Spectra under Resonant Two-Photon Excitation of the Biexciton: Additional Measurements for Different Pulse Durations}
\label{app:2dColormapAdditionalMeasurements}
\begin{figure}
    \centering
    \includegraphics{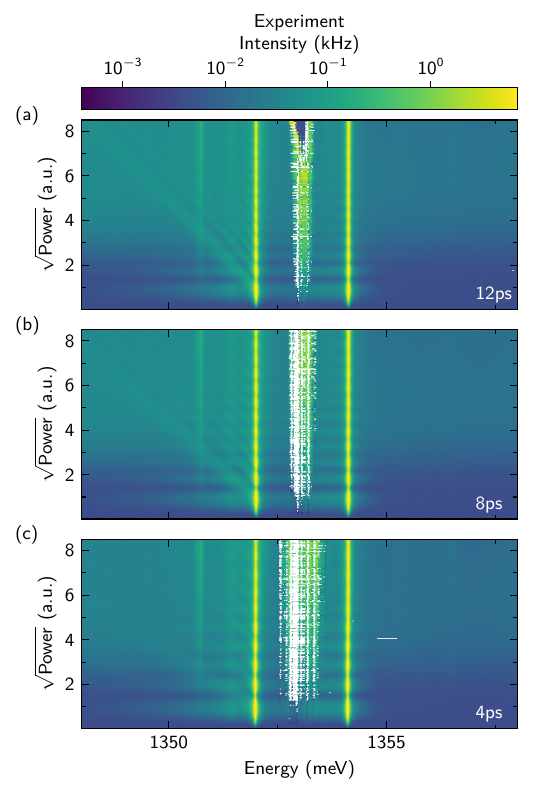}
    \caption{Logarithmically plotted intensity of the measured power-dependent emission spectra for pulse durations of (a) \SI{12}{\ps}, (b) \SI{8}{\ps} and (c) \SI{4}{\ps}.
    }
    \label{fig:power-dependentColormap}
\end{figure}
\cref{fig:power-dependentColormap} shows additional plots displaying the logarithmic emission intensity versus both the emission energy and the square root of the excitation power for pulse durations of \SI{12}{\ps}, \SI{8}{\ps} and \SI{4}{\ps}, respectively.
As discussed in the main manuscript, the observed intensity oscillations of the biexciton and exciton emission lines confirm the coherent driving of the four-level system under resonant two-photon excitation. 
The phonon sideband appears red-detuned from the biexciton and exciton emission energies at odd multiples of $\mathrm{\pi}$ in the Rabi rotations. In addition, with increasing excitation power, multiple side peaks evolve red-detuned from the biexciton transition energy.
The pulse length dependent features observed by comparing the emission spectra of \SI{12}{\ps}, \SI{8}{\ps} and \SI{4}{\ps} agree well with our observations in the main manuscript. 
As the pulse duration decreases, the emission intensity of the side peaks decreases, the dressed part of the spectrum broadens, and the number of $\mathrm{\pi}$ rotations and thus the number of side peaks within a given power range decreases.

\section{Two-Photon Pulse Area Conversion into the Power Domain}
\label{app:pulsearea}
In the main manuscript, the emission spectra and Rabi rotations were investigated as a function of the excitation power rather than the effective pulse area. 
In order to investigate the nonlinear pulse area of resonant two-photon excitation of the biexciton from an experimental point of view, we establish the dependence of the theoretical description of the effective pulse area for low driving introduced by Stufler \textit{et al.} \cite{stufler2006TPE} on the power domain.

Given the relationship between the electric field of a Gaussian shaped laser pulse and its intensity, the average power can be described by
\begin{equation}
        \mathrm{P_{Avg} \propto \int_{-\infty}^{\infty} dt \abs{E_0}^2 e^{-4\ln{2}\frac{t^2}{\tau^2}} \propto E_0^2 \tau },
\end{equation}
where $\mathrm{\tau}$ is the intensity full width at half maximum, which in the main manuscript is defined as the pulse duration. 
Let us now introduce the effective pulse area $\mathrm{\Lambda(\infty)}$ for the two-photon resonant excitation of the biexciton state. 
For low driving ($\mathrm{\Theta \ll \frac{\tau_i E_b}{\hbar}}$) the effective pulse area is theoretically described by Stufler \textit{et al.} \cite{stufler2006TPE} as
\begin{equation}
    \mathrm{\Lambda(\infty) \approx \frac{4\hbar \, arcosh \sqrt{2}}{\pi^2 E_b \tau}\Theta^2}
    \,,
\end{equation}
where $\mathrm{E_b}$ is the biexciton binding energy and $\mathrm{\Theta}$ is the pulse area of a resonantly driven two-level system, which is defined as
\begin{equation}
    \mathrm{\Theta = \frac{\mu_{12}}{\hbar} \int_{-\infty}^{\infty} dt E_0 e^{-2\ln{2}\frac{t^2}{\tau^2}}}
    \,.
\end{equation}
If we now insert $\mathrm{\Theta}$ into the effective pulse area of the studied four-level system, and if we consider $\mathrm{E_b}$ to be constant for a single QD, we obtain the following relationship between the effective pulse area and the excitation power:
\begin{equation}
    \mathrm{\Lambda(\infty) \propto E_0^2 \tau \proptoinverse  P_{Avg}}
    \,.
\end{equation}
Up to approximately $\mathrm{1\pi}$ in the main manuscript and for typical binding energies of \SI{2}{\meV} \cite{finley2001chargedExcitonComplexes}, the condition $\mathrm{\Theta \ll \frac{\tau E_b}{\hbar}}$ of low driving holds true. In this regime, the effective pulse area of the TPE is independent of the pulse duration and scales linearly with the average excitation laser power.

\section{Emission Spectra under Detuned Two-Photon Excitation of the Biexciton: Additional Measurements for Different Pulse Areas}
\label{app:detuningAdditionalMeasurements}
\begin{figure}
    \centering
    \includegraphics{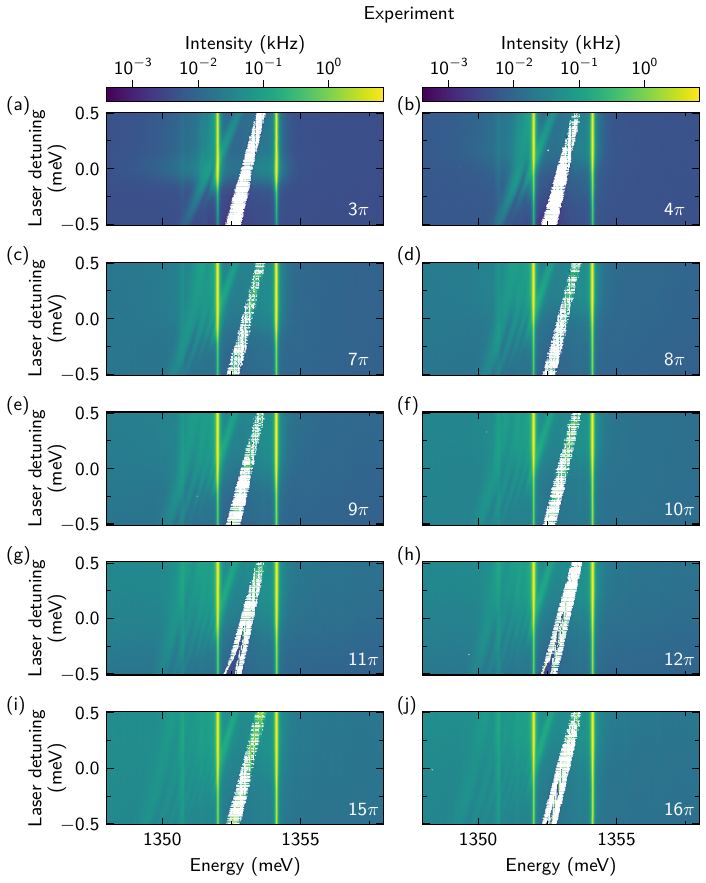}
    \caption{
     Measured emission spectra under detuned pulsed excitation by $\mathrm{\pm \SI{0.51}{\milli\eV}}$ over the TPE resonance. An excitation pulse with a pulse duration of \SI{14}{\ps} and pulse areas of (a) $\mathrm{3\pi}$, (b) $\mathrm{4\pi}$, (c) $\mathrm{7\pi}$, (d) $\mathrm{8\pi}$, (e) $\mathrm{9\pi}$, (f) $\mathrm{10\pi}$, (g) $\mathrm{11\pi}$, (h) $\mathrm{12\pi}$, (i) $\mathrm{15\pi}$, (j) $\mathrm{16\pi}$ were chosen.
    }
    \label{fig:detunedColormap}
\end{figure}
To confirm that the characteristics observed in the main manuscript for pulsed excitation with finite detuning remain consistent across different excitation powers, we present additional emission spectra in \cref{fig:detunedColormap}.
These spectra were recorded under pulsed excitation, where a \SI{14}{\ps} pulse was scanned by $\mathrm{\pm \SI{0.51}{\milli\eV}}$ across the TPE resonance. The measurements were taken at constant powers corresponding to pulse areas of $\mathrm{3\pi, 4\pi, 7\pi, 8\pi, 9\pi, 10\pi, 11\pi, 12\pi, 15\pi, 16\pi}$ respectively.
The detuning dependent shift of the side peaks is consistent with the observations made in the main manuscript. 
For $\mathrm{\Delta<0}$, the emission energy of the side peaks follows the laser detuning, while for $\mathrm{\Delta >0}$, the side peaks asymptotically approach the main biexciton emission line, except for the innermost side peak, which continues to follow the laser detuning.
Comparing panels (a--j) of \cref{fig:detunedColormap}, it can be seen that as the pulse area changes, a pulse area specific number of side peaks appear, also in agreement with the observations in the main part of the paper.
It can also be observed that the background emission intensity increases with increasing pulse area.
This background is sample specific and occurs when the QD is subject to very strong electric fields, e.g. upon excitation with multiples of $\mathrm{\pi}$. 
Notably, the background emission is absent when the QD is tuned out of resonance but the laser is still shining on the sample. 
It follows that the on/off measurement technique does not take this background contribution into account. 
The physical origin of this background may be related to background luminescence from impurities and charge traps within the semiconductor host matrix.

\section{Time-Dependent Measurements}
\label{app:timeDependentMeasurements}

\subsection{Biexciton and Exciton Lifetime Analysis}
\label{subapp:Lifetime_XX_X}
\begin{figure}
    \centering
    \includegraphics{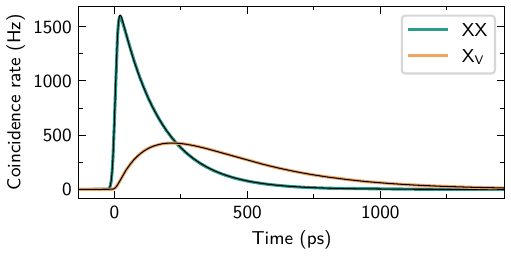}
    \caption{
    Time-resolved measurements of the biexciton-exciton cascade with a biexciton lifetime (teal) of \SI{157}{\ps} and an exciton lifetime (orange) of \SI{295}{\ps}. The black lines represent the fitted curves.
    }
    \label{fig:lifetimeX&XX}
\end{figure}
The time evolution of the biexciton-exciton cascade is displayed in \cref{fig:lifetimeX&XX}. At $\mathrm{t= \SI{0}{\ps}}$, the laser pulse which excites the four-level system from the ground state to the biexciton state has its maximum field intensity. Over the course of the pulse, the biexciton state is populated, whose decay results in a uniform build-up of the exciton state population.
A biexciton lifetime of \SI{157}{\ps} and an exciton lifetime of \SI{295}{\ps} can be extracted. These are long compared to the emission decay times of the side peaks and the excitation pulse length, which was in the range of \SI{4}{\ps} to \SI{14}{\ps} in the main manuscript. This allows us to study the dressed states in the dynamic regime, which requires a pulse duration shorter than the lifetime of the studied system \cite{KRzazewski_1984, finley2004starkShift, gustin18pulsedExcitation}.

To extract the biexciton lifetime $\mathrm{\tau_{XX}}$ (teal) from the time-resolved measurement shown in \cref{fig:lifetimeX&XX}, we assume an instantaneous population of the biexciton state, i.e., $\mathrm{I_{XX}(0)=1}$, and fit the data using a mono-exponential decay function $\mathrm{I_{XX}(t)=e^{-\frac{t}{\tau_{XX}}}}$. 
The build-up of the biexciton population during the excitation pulse is neglected, as its timescale is shorter than the temporal resolution of our measurement setup.
To determine the exciton lifetime $\mathrm{\tau_X}$, it must be taken into account that the exciton state is populated via the cascaded decay of the biexciton.
This process is described by the coupled rate equation:
\begin{equation}
    \mathrm{ \frac{dI_{X,TPE}(t)}{dt} = -\frac{I_{X,TPE}(t)}{\tau_X} + \frac{I_{XX}(t)}{\tau_{XX}} }
    \,,
    \label{eq:coupledRateEq}
\end{equation}
where an analytic solution can be obtained by assuming $\mathrm{I_X(0)=0}$ \cite{bateman1910}, yielding a bi-exponential expression:
\begin{equation}
    \mathrm{ I_{X,TPE}(t) = -\frac{\tau_X}{\tau_{XX}-\tau_X}e^{-\frac{t}{\tau_X}} + \frac{\tau_X}{\tau_{XX}-\tau_X}e^{-\frac{t}{\tau_{XX}}} }
    \,.
\end{equation}
In \cref{eq:coupledRateEq}, $\mathrm{I_{XX}(t)}$ and $\mathrm{I_{X,TPE}(t)}$ denote the time-dependent populations of the biexciton and exciton states, respectively.

To account for the timing jitter of the SNSPDs used in our experiment, the instrument response function (IRF) is incorporated by convolving the decay functions with a Gaussian distribution, leaving the standard deviation as a free fit parameter yielding $\mathrm{\sigma \approx \SI{9}{\ps}}$.
This results in the following fit function for the biexciton decay  with the population of the biexciton occurring at the time $\mathrm{t_0}$:
\begin{equation}
    \mathrm{ I_{XX}(t) = \biggl( \mathcal{H}(t-t_0)A e^{-\frac{t-t_0}{\tau_{XX}}} \biggr) * \biggl(\frac{1}{\sqrt{2\pi}\sigma} e^{-\frac{t^2}{2\sigma^2}}\biggr) }
    \,,
    \label{eq:Convolution-I_XX(t)}
\end{equation}
and for the exciton decay:
\begin{equation}
    \mathrm{ I_{X,TPE}(t)= \biggl( \mathcal{H}(t-t_0)A\biggl( -\frac{\tau_X}{\tau_{XX}-\tau_X}e^{-\frac{t-t_0}{\tau_X}} + \frac{\tau_X}{\tau_{XX}-\tau_X} e^{-\frac{t-t_0}{\tau_{XX}}} \biggr) \biggr) * \biggl( \frac{1}{\sqrt{2\pi}\sigma} e^{-\frac{t^2}{2\sigma^2}} \biggr) }
    \,,
    \label{eq:convolution-I_XTPE(t)}
\end{equation}
where $\mathcal{H}$ is the Heaviside function, $\mathrm{A}$ is a scaling factor and $\mathrm{t_0}$ is the offset in time of the acquired histogram.
An analytic solution to \cref{eq:Convolution-I_XX(t)} and \cref{eq:convolution-I_XTPE(t)} can be derived by a twofold substitution with $\mathrm{u=\tau- \bigl(t-\frac{\sigma^2}{\tau_{XX(X)}} \bigr)}$ and $\mathrm{v=\frac{u}{\sqrt{2}\sigma}}$, yielding:
\begin{equation}
    \mathrm{I_{XX}(t) = A e^{\frac{1}{2}\bigl(\frac{\sigma}{\tau_{XX}}\bigr)^2 -\frac{(t-t_0)}{\tau_{XX}}} \biggl[ erf\biggl( \frac{t-t_0}{\sqrt{2}\sigma}  - \frac{\sigma}{\sqrt{2}\tau_{XX}}\biggr) +1 \biggr] }
    \,,
    \label{eq:fitfunction_XX}
\end{equation}
and 
\begin{equation}
\begin{split}
    \mathrm{ I_{X,TPE}(t) = -\frac{\tau_X}{\tau_{XX}-\tau_X}A \biggl[ } & \mathrm{ e^{\frac{1}{2} \bigl( \frac{\sigma}{\tau_X} \bigr)^2 - \frac{t-t_0}{\tau_X}} \biggl( erf\biggl( \frac{t-t_0}{\sqrt{2}\sigma} - \frac{\sigma}{\sqrt{2}\tau_X} \biggr) +1 \biggr) } \\
    & \mathrm{ -e^{\frac{1}{2}\bigl( \frac{\sigma}{\tau_{XX}} \bigr)^2 - \frac{t-t_0}{\tau_{XX}}} \biggl( erf\biggl( \frac{t-t_0}{\sqrt{2}\sigma} - \frac{\sigma}{\sqrt{2}\tau_{XX}} \biggr) +1 \biggr) \biggr] }
    \,.
\end{split}
\label{eq:fitfunction_XTPE}
\end{equation}
We use \cref{eq:fitfunction_XX} and \cref{eq:fitfunction_XTPE} to fit the time-dependent photoluminescence measurements in \cref{fig:lifetimeX&XX}.

\subsection{Theoretical calculation of time-dependent occurrence of the sidebands}
\label{subapp:theoryTimeDependentOccurrence}
An approximate analytical model derived in Ref.~\cite{moelbjerg2012resonance} can be used for the two-level system to derive the points in time, at which the sidebands occur in the time-dependent spectrum. We will briefly recap their findings and highlight the steps needed to apply the model to driving of the two-photon resonant ground-state to biexciton transition.

For a decoherence-free two-level system driven by an external laser field, the population $\mathrm{X}$ and the coherence $\mathrm{\rho_{GX}}$ is given by 
\begin{equation}
    \mathrm{X(t)=\langle\ket{X}\bra{X}\rangle(t)=\sin(\phi(t)/2), \quad \rho_{GX}(t)=\langle\ket{G}\bra{X}\rangle(t)=\frac{i}{2}\sin\left[\phi(t)\right]} \label{eq:x_p}
\end{equation} are fully described by the Rabi phase $\mathrm{\phi(t) = \int_{-\infty}^{t}\Omega(t')dt'}$, where $\mathrm{\Omega(t)}$ is the Rabi frequency induced by the resonant laser pulse. For $\mathrm{t\rightarrow\infty}$, the Rabi phase corresponds to the pulse area.

The criterion that can be found for the times where the side peaks emerge depends on the Rabi phase as well as the Rabi frequency and reads \cite{moelbjerg2012resonance}
\begin{equation}
    \mathrm{\phi(t) - \phi(-t) - 2\Omega(t)t = (2n+\frac{1}{2})\pi}
    \,.
    \label{eq:t_sidebands}
\end{equation}
This assumes that the pulse is centered around $\mathrm{t=0}$ and $\mathrm{n}\in\mathbb{N}_0$. Using the solutions $\mathrm{t_n}$ to \cref{eq:t_sidebands} can additionally be used to calculate the spectral position of the sidebands, by considering the splitting of the dressed states at these points in time.

The challenge in doing a similar analysis for other transitions of the system, like excitation of the biexciton $\mathrm{XX}$, lies in finding expressions for $\mathrm{\phi(t)}$ that accurately describe the system dynamics in a form similar to the two-level system given in \cref{eq:x_p}. As described in \cref{app:pulsearea} no exact analytical expressions exist for the case of the ground-state to biexciton transition. The approximate analytic expressions do usually not lead to accurate results in this case, but a numerical fit of $\mathrm{\phi(t)}$ based on the decoherence-free time dynamics of the system can be done to obtain accurate results, as visible in \cref{fig:timeDependentIRS} in the following section.

\subsection{Time-Dependent Simulation of the Side Peak Emission with and without Instrument Response Function}
\label{subapp:timeDependentIRF}
\begin{figure}
    \centering
    \includegraphics{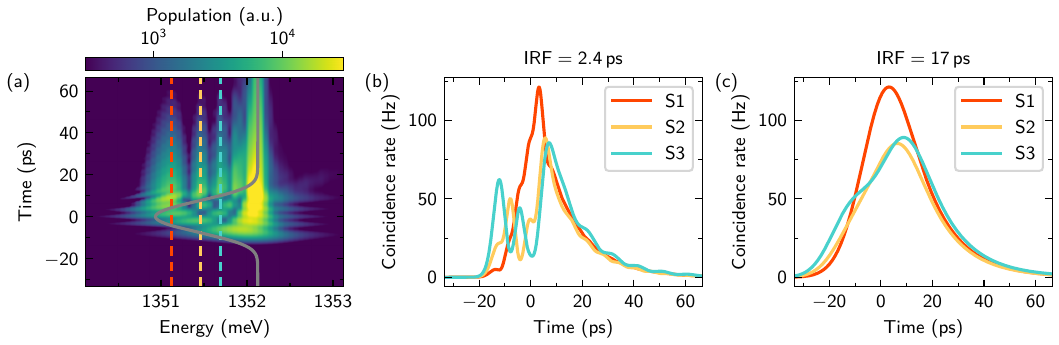}
    \caption{
    Simulations of time-resolved emission under different detector conditions.
    (a) Simulated spectrum assuming an ideal detector with perfect temporal resolution.
    (b) Time-resolved photon emission of $\mathrm{S_1}$, $\mathrm{S_2}$ and $\mathrm{S_3}$ using an almost ideal detector characterized by an IRF of \SI{2.4}{\ps}.
    (c) Time-resolved photon emission calculated using detector parameters similar to those of the experimental setup described in the main manuscript, i.e., with an IRF of \SI{17}{\ps}.
    }
    \label{fig:timeDependentIRS}
\end{figure}
In the main manuscript, the measured time-dependent emissions of the side peaks $\mathrm{S_1}$, $\mathrm{S_2}$ and $\mathrm{S_3}$ were found to have a short temporal width comparable to the excitation pulse duration and a temporal profile similar to that of a Gaussian pulse.
Due to the finite picosecond temporal resolution of our detectors, and in general of all currently commercially available detectors, the temporal width and shape of the side peaks observed in the experiment carry only limited information about the actual temporal emission properties and are mainly determined by the timing jitter of the detector.
To determine the impact of detector properties on the measured time-resolved emission characteristics of the side peaks, we study the calculated time-resolved photon emission under varying detector conditions. In particular, we compare the time-resolved emission obtained using an ideal detector with that obtained using a detector with the same temporal resolution as that used in the experimental setup described in the main manuscript.

In contrast to the time-dependent, time-integrated emission spectrum presented in the main manuscript, we now study a time-dependent emission spectrum calculated using the sensor formalism introduced by Elena del Valle \textit{et al.} \cite{delValle2012sensor}. 
For this simulation, detectors with a linewidth of \SI{0.05}{\milli\eV} are simulated.
The resulting spectrum is shown in \cref{fig:timeDependentIRS}(a), where the population of a two-level sensor coupled to the four-level QD system is plotted as a function of time and emission energy.
Consistent with the observations in the main manuscript, we see that as the excitation pulse (gray) arrives and evolves over time, multiple side peaks emerge from the main biexciton emission line. These peaks shift towards lower energies and interfere with each other until, at certain times, each peak stabilizes at a specific energy and reaches a maximum before the population of the two-level sensor system decays. 
However, unlike in the time-dependent, time-integrated spectrum, the sensor population in \cref{fig:timeDependentIRS}(a) decays to zero after the excitation pulse has passed.

This sensor formalism used to describe the time-dependent emission spectrum allows us to obtain an equivalent to the time-resolved photon emission measurement presented in the main manuscript. 
Therefore, we frequency filter the side peaks $\mathrm{S_1}$ (red dashed line), $\mathrm{S_2}$ (yellow dashed line) and $\mathrm{S_3}$ (light blue dashed line) at maximum absorption (\cref{fig:timeDependentIRS}(a)) and obtain the simulated time-resolved emissions of the side peaks as shown in \cref{fig:timeDependentIRS}(b) and \cref{fig:timeDependentIRS}(c).
Both figures show the time-dependent emission characteristics of $\mathrm{S_1}$, $\mathrm{S_2}$ and $\mathrm{S_3}$, but differ in the parameter $\mathrm{\sigma}$, which, as defined in \cref{eq:Convolution-I_XX(t)}, represents the standard deviation of a Gaussian distribution used to model the detector instrument response function in the simulations.
In \cref{fig:timeDependentIRS}(b) and \cref{fig:timeDependentIRS}(c), values of $\mathrm{\sigma = \SI{1}{\ps}}$ and $\mathrm{\sigma = \SI{7.2}{\ps}}$ were used, corresponding to detector timing jitter with a FWHM of \SI{2.4}{\ps} and \SI{17}{\ps}, respectively. 
$\mathrm{\sigma = \SI{1}{\ps}}$ represents an almost ideal detector, whereas $\mathrm{\sigma = \SI{7.2}{\ps}}$ was chosen to match the temporal resolution of the detector used in the time-resolved experimental measurements presented in the main manuscript.

\cref{fig:timeDependentIRS}(b) shows the time-dependent photon emission as it would be detected by an almost ideal detector. 
Up to $\mathrm{t=\SI{0}{\ps}}$, when the pulse reaches its maximum field amplitude, interference fringes from $\mathrm{S_1}$, $\mathrm{S_2}$ and $\mathrm{S_3}$ are clearly visible. These fringes persist as the pulse evolves.
As the side peaks evolve, their coincidence rates increase and reach a maximum at the moment they reach their final spectral position, as shown in \cref{fig:timeDependentIRS}(a). 
After this point, the emission rates of the side peaks decay to zero.
In contrast, \cref{fig:timeDependentIRS}(c) shows the time-resolved photon emission simulated using a detector with an IRF matching the temporal resolution of the detector used in the main manuscript. In this case, no interference fringes are observed. 
Instead, the temporal emission profile appears smeared out, with an overall broader temporal width and a more Gaussian-shaped temporal profile.
This comparison shows that the temporal width and shape of the detected photon emission from the side peaks is primarily determined by the temporal resolution of the detector.
Nevertheless, the key characteristic presented in the main manuscript, the temporal ordering of the side peak emission maxima in a spectrum ($\mathrm{S_1}$, followed by $\mathrm{S_2}$, followed by $\mathrm{S_3}$), remains evident in both \cref{fig:timeDependentIRS}(b) and \cref{fig:timeDependentIRS}(c).

\end{document}